\documentclass[usenatbib,usegraphicx,twocolumn]{mn2e}
\usepackage[english]{babel} 
\usepackage{color}

\def\HI{H~{\sc i}}

\def\pasj{PASJ}

\def\mnras{MNRAS}

\def\aap{A \& A}
\def\aaps{Astronomy and Astrophysics Supplement Series}
\def\apj{ApJ}
\def\apjs{ApJS}
\def\apjl{ApJL}

\def\V2{V_2}
\def\V2ij{V_{2ij}}
\def\S{{\mathcal S}}
\def\V{\mathcal{V}}

\def\la{\left\langle}

\def\lsim{~\rlap{$<$}{\lower 1.0ex\hbox{$\sim$}}}

\def\gsim{~\rlap{$>$}{\lower 1.0ex\hbox{$\sim$}}}

\usepackage{graphics}
\usepackage{color}
\usepackage{amsmath}
\usepackage{amssymb}
\usepackage{psfrag}
\usepackage{subfig}
\usepackage{graphicx}

\newsavebox{\measurebox}
\begin{document}
\date {} \title[Turbulence in warm and cold neutral medium] {Turbulent
  power spectrum in warm and cold neutral medium using the Galactic
  H~{\sc i} 21 cm emission} \author[S. Choudhuri et al.]{Samir
  Choudhuri$^{1}$\thanks{Email:samir@ncra.tifr.res.in} and Nirupam
  Roy$^{2}$\\ $^{1}$ National Centre For Radio Astrophysics, Post Bag
  3, Ganeshkhind, Pune 411 007, India\\ $^{2}$ Department of Physics,
  Indian Institute of Science, Bangalore 560012, India}

\maketitle

\begin{abstract}
Small-scale fluctuations of different tracers of the interstellar the
medium can be used to study the nature of turbulence in astrophysical
scales. Of these, the ``continuum'' emission traces the fluctuations
integrated along the line of sight whereas, the spectral line tracers
give the information along different velocity channels as
well. Recently, \citet{miville16} have measured the intensity
fluctuation power spectrum of the continuum dust emission, and found a
power law behaviour with a power law index of $-2.9 \pm 0.1$ for a
region of our Galaxy. Here, we study the same region using
high-velocity resolution 21-cm emission from the diffuse neutral
medium, and estimate the power spectrum at different spectral
channels. The measured 21-cm power spectrum also follows a power law,
however, we see a significant variation in the power law index with
velocity. The value of the power-law index estimated from the
integrated map for different components are quite different which is
indicative of the different nature of turbulence depending on
temperature, density and ionization fraction. We also measure the
power spectra after smoothing the 21 cm emission to velocity
resolution ranging from $1.03$ to $13.39~{\rm km~s^{-1}}$, but the
power spectrum remains unchanged within the error bar. This indicates
that the observed fluctuations are dominantly due to density
fluctuations, and we can only constrain the power-law index of
  velocity structure function of $0.0 \pm 1.1$ which is consistent
  with the predicted Kolmogorov turbulence $(\gamma=2/3)$ and also with a
    shock-dominated medium $(\gamma=1.0)$.
\end{abstract} 


\section{Introduction}
Study of the Galactic neutral hydrogen (\HI) 21 cm emission can reveal
the kinematic and morphological properties of the interstellar medium
(ISM) of our Galaxy. Several all-sky surveys were performed at
different angular scales to probe the distribution of \HI~within the
Galaxy. For example, the Leiden/Argentine/Bonn (LAB) Survey
\citep{kalberla05} is one of the most sensitive Milky Way \HI~surveys
which has extensive coverage both spatially and
kinematically. Recently, \HI 4$\pi$ survey \citep{hi4pi} supersedes
the LAB survey with a higher angular resolution. Neutral atomic ISM
has structures over a wide range of scales. Early observations
indicated the presence of cold \HI~in the form of ``clouds'' of size
few parsec \citep{field69,field73}. Further observations with high
resolution and sensitivity show the structure of even smaller scales
of the order of sub-pc \citep{crovisier85,kalberla85}. These
structures are believed to be generated by the compressible turbulence
in the ISM. \citet{dieter76} have reported variation in the optical
depth of interstellar \HI~on 70 AU scale in front of a quasar 3C~147
using Very Long Baseline Interferometry (VLBI)
observations. Similarly, there are multi-epoch pulsar observations
which reveal the opacity variations on several AU scales
\citep{deshpande92,frail94,davis96,johnston03,stanimirovic10}.

Statistical measurements of the power spectrum, $P({\rm k})$ is a
powerful technique to probe and quantify the small-scale structures in
the ISM. Observations of \HI~ 21 cm emission have shown power law
power spectra on pc scales
\citep{crovisier83,green93,dickey01}. \citet{roy10} and
\citet{deshpandeal00} have measured power spectra to quantify the
fluctuations even at smaller scales ($\sim$AU scale) for different
parts of the Galaxy. The structure function of the opacity
fluctuations have also been used to study the tiny scale
fluctuations. For example, \citet{roy12} and \citet{dutta14} have used
\HI~absorption to measure the structure-function of the opacity
fluctuations down to a few tens of AU scale, and showed it to have a
power law scaling. Even if the details are not completely understood,
analytical and numerical works
\citep{deshpande00,nagashima06,vazquez06,hennebelle07} suggest
turbulence as the key mechanism for generating these small-scale
structures.

From an observational point of view, the 21 cm emission power spectrum
could be modified due to the turbulent velocity fluctuations in the
field.  \citet{lazarian00} showed that velocity fluctuations make the
estimated power spectra shallower, provided that the velocity width of
the individual spectral channel is much smaller than the turbulent
dispersion of the \HI~ gas. To test this, \citet{deshpandeal00}
estimated the \HI~power spectrum towards the directions of Cas~A and
Cygnus~A with and without velocity averaging, but they did not find
any change in the slope of the measured power spectrum. It indicates
that the contribution of the velocity fluctuations is less significant
in the measured power spectrum for those directions. Later,
\citet{roy10} have also measured the opacity power spectrum in the
direction of Cas~A with better velocity resolution, and they also have
not found any change in the power law index due to velocity averaging.

In this paper, we study the \HI~21 cm power spectrum in the direction
of Galactic coordinate ({\it l,b})=$(198^{\circ},32^{\circ}$).  For
this direction, \citet{miville16} have used the {\it Planck} radiance
map at $857 {\rm GHz}$, the MegaCam g band map and the WISE $12\mu
{\rm m}$ emission map to estimate the dust continuum emission power
spectrum. Combining these three observations, they measured the
spatial power spectrum over a large range of scales ranging from
$0.01~{\rm pc}$ to $50~{\rm pc}$. They found a single power law with
index $-2.9\pm0.1$, consistent with the density field power spectrum
of turbulent neutral ISM. They have given some estimate of the scale,
less than $0.01~{\rm pc}$, at which the turbulent energy dissipates.
Here, we use the spectral information of \HI~ 21 cm emission for power
spectrum estimation for narrow velocity channels. Our aims here are:
(a) to measure the variation of turbulence power spectra in different
velocity channels, and (b) to quantify the fluctuations of the
velocity structures, if present, by estimating power spectrum for
different velocity resolution.  We use the high-velocity resolution
LAB survey \citep{kalberla05} data for this purpose.

A brief outline of the paper follows. Data analysis is presented in
Section 2. In Section 3, we present the results of the power spectrum
measurement. Finally, we summarize and conclude in Section 4.

\section{Data Analysis}
\label{data}
LAB survey presents all-sky 21 cm emission spectra by combining the
Leiden/Dwingeloo Survey \citep{hartmann97} and the Instituto Argentino
de RadioastronomÃ­a Survey \citep{arnal00,bajaja05} data. The angular
resolution of this survey is around half a degree.  The velocity range
covers $-450~{\rm km~s^{-1}}$ to $400~{\rm km~s^{-1}}$ with a channel
separation $1.03~{\rm km~s^{-1}}$. The rms brightness temperature of
the survey is $0.07-0.09~{\rm K}$.

\begin{figure}
\begin{center}
\includegraphics[width=80mm,angle=0]{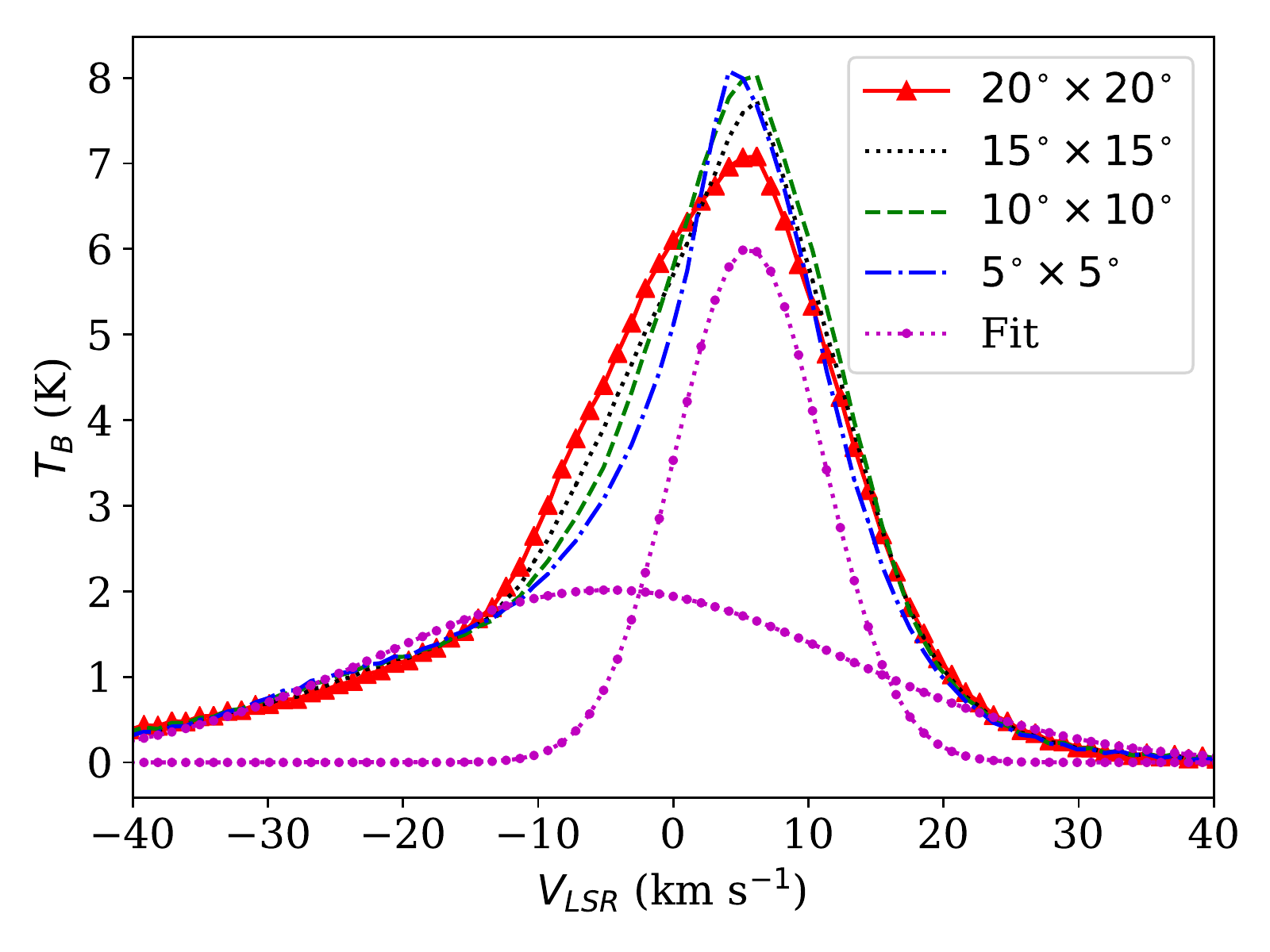}
\caption{The \HI~21 cm spectra averaged over different spatial dimensions centred on Galactic coordinate $(198^{\circ},32^{\circ})$. The magenta circles show the two components Gaussian fits to the spatially averaged spectra from the central $5^{\circ}\times5^{\circ}$ region.}
\label{fig:fig1}
\end{center}
\end{figure}

We use the NRAO Astronomical Image Processing System (AIPS) package
for further analysis of the LAB survey data.  We select a region
centred at Galactic coordinate $(198^{\circ},32^{\circ})$ from the LAB
survey for our purpose. Figure \ref{fig:fig1} shows the spatially
average spectra over a velocity range of $-40~{\rm km~s^{-1}}$ to
$+40~{\rm km~s^{-1}}$. The red solid line with triangles shows the
spectra averaged over angular dimensions $20^{\circ}\times20^{\circ}$.
The peak value of this spectra is around $8~{\rm K}$ at a velocity
$+4~{\rm km~s^{-1}}$. The black dotted, green dashed, and blue
dash-dot lines show different spectra averaged over an image of
dimensions $15^{\circ}\times15^{\circ}$, $10^{\circ}\times10^{\circ}$
and $5^{\circ}\times5^{\circ}$ respectively. Please note that the
spectra are not symmetric along the positive and negative velocity
ranges with respect to the peak. There is some wide emission component
in the negative velocity region whereas the spectra fall relatively
more sharply in positive velocity range. The best two-component
Gaussian fit (magenta circles) of the averaged spectra from the
central $5^{\circ}\times5^{\circ}$ region indicates the presence of
two distinct phases: (a) the cold phase (Cold Neutral Medium; CNM)
with smaller width ($\sigma_v = 5.4~{\rm km~s^{-1}}$), and (b) the
warm phase (Warm Neutral Medium; WNM) with significantly larger width
($\sigma_v=17.3~{\rm km~s^{-1}}$). In Figure \ref{fig:fig1} we see
that the negative velocity range is dominated by the WNM whereas the
positive velocity range has more contribution from the CNM. We
  divide the velocity range $-30$ to $+20~{\rm km~s^{-1}}$ with four
  different regions. The velocity range from $-30$ to
  $-15~{\rm km~s^{-1}}$ can be interpreted as an intermediate-velocity gas component
  (IVC) as shown in Figure 4 of \citet{miville16}. The velocity range
  from $-14$ to $-5~{\rm km~s^{-1}}$ is mainly dominated by the WNM of
  the Galaxy disk. The CNM is dominated in the velocity range from
  $-4$ to $+15~{\rm km~s^{-1}}$. The velocity range $\ge+16~{\rm
    km~s^{-1}}$ is again dominated by the WNM, but the relative
  amplitude is smaller here as compared to the WNM in the negative
  velocity range. In the next section, we use the
  $20^{\circ}\times20^{\circ}$ image with velocity ranges from $-30$
  to $+20~{\rm km~s^{-1}}$ to study the statistical properties of
  these components in terms of the power spectrum. We use this
velocity span because the power spectrum of brightness temperature
fluctuations becomes gradually more noise dominated.

We also measure the power spectrum after averaging the velocity
channels of different width.  For this purpose, we smooth the cube
with a Boxcar function along the frequency by changing the width of
the function. We use ``n''-pixel (n=3,5,7,...) boxcar smoothing
  function centred on each frequncy channel to make a lower resolution
  cube.  We first use AIPS task TRANS and then XSMTH for smoothing.
  The channel separation of LAB data is $1.03~{\rm km~s^{-1}}$, whereas
  the resolution of the LAB survey is $1.25~{\rm km~s^{-1}}$
  (\citealt{kalberla05}, Table 1). The channel resolution for this
  velocity smoothing will be slightly more than ``n'' times channel
  separation due to this correlation. While considering the variation
  of power spectra after smoothing, we consider channels separated by
  ``n'' pixels to consider only independent measurements. We use velocity
  averaged cubes for the power spectrum measurement in Section
  \ref{psanalysis}.

\section{Power Spectrum Analysis}
\label{psanalysis}
\begin{figure}
\begin{center}
\includegraphics[width=80mm,angle=0]{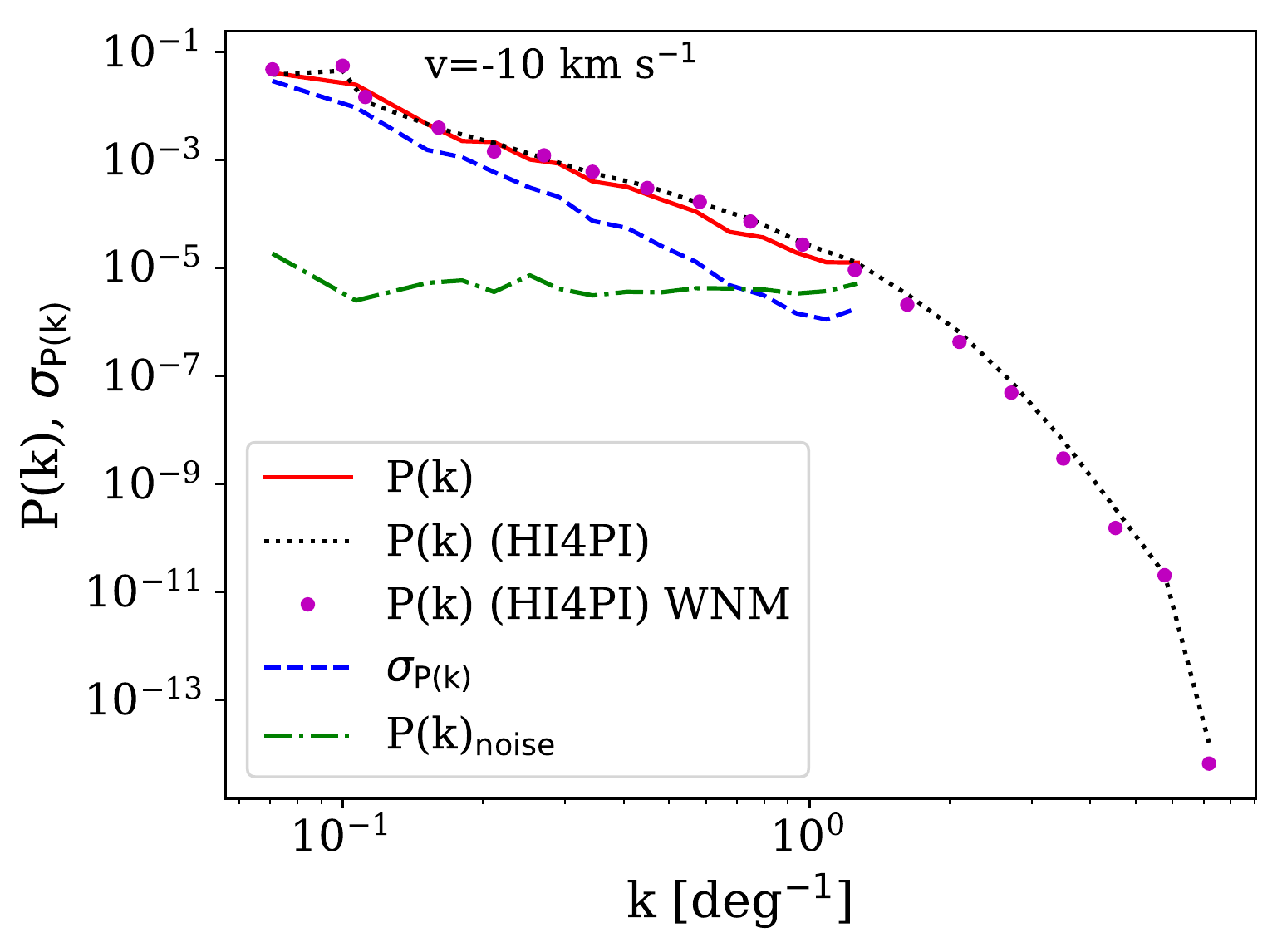}
\caption{The estimated power spectrum, $P({\rm k})$, for
  $v_{LSR}=-10~{\rm km~s^{-1}}$ which is mainly dominated by WNM (red solid line). For
  comparison, we also show the estimated $P({\rm k})$ for a line free
  channel which is mainly dominated by the system noise (green
  dash-dot line). Note that the noise power spectrum is much lower
  than the 21 cm signal power spectrum. The measured $P({\rm k})$ for
  $v_{LSR}=-10~{\rm km~s^{-1}}$ using \HI 4$\pi$ Survey is also shown by
  black dotted line. The blue dash line shows the expected cosmic
  variance for this channel assuming the signal to be Gaussian random
  in nature. The magenta points show the velocity integrated
power spectrum in the WNM dominated region $(-14 < v_{LSR} <-5 ~{\rm
    km~s^{-1}})$ for \HI 4$\pi$ data which is remain same as for the single channel power specturm.}
\label{fig:fig1a}
\end{center}
\end{figure}
We use the Fastest Fourier Transform in the West (FFTW;
\citealt{frigo05}) to convert the image in the Fourier domain and then
to estimate the power spectrum at different Fourier modes. To
  avoid ringing due to sharp cut-off at the edge, we multiply the
  image with a cosine function.  Also, we average azimuthally to
  increase the signal to noise (SNR) in the final power spectrum
  estimation. \citealt{kalberla16,kalberla17} have studied the
  anisotropy in the power spectrum for HI gas distribution. But, we
  have not seen any significant anisotropies in the measured power
  spectrum for this field. So, we have considered only the azimuthally
  averaged power spectrum in the rest of the paper. As mentioned
earlier, we use an image of dimensions $20^{\circ}\times20^{\circ}$
with angular resolution $0.5^{\circ}$ for this purpose. We divide the
whole k-range $(0.07\le{\rm k}\le1.2~{\rm deg}^{-1})$ into $20$
equally spaced logarithmic bins to increase the SNR in the estimated
$P({\rm k})$. Assuming that the physical distance of the \HI~clouds is
$200~{\rm pc}$ \citep{miville16}, here we probe the fluctuations of
length scale ranging from $3~{\rm pc}$ to $50~{\rm pc}$.

In Figure \ref{fig:fig1a} we show the measured $P({\rm k})$ as a
function of ${\rm k}$ (in ${\rm deg^{-1}}$) at LSR velocity
  $v_{LSR}=-10~{\rm km~s^{-1}}$ which is mainly dominated by the WNM
(red solid line). The green dashed-dot line shows the measured
  $P({\rm k})$ due to the system noise for a line free channel.  We
see that the noise power spectrum is roughly constant at different
Fourier modes and the magnitude is much lower than the line channel
power spectrum. Assuming the signal to be Gaussian in nature, we
estimate the expected cosmic variance in the power spectrum
measurement which is shown by the blue dashed line. Although the
assumption may not be true for Galactic \HI~ signal, we have used this
approximation for error estimation in the rest of the paper because we
don't otherwise have any analytical formula for estimating the error
from a single realization of the signal. In Figure \ref{fig:fig1a}, we
also show the estimated $P({\rm k})$ at $v_{LSR}=-10~{\rm km~s^{-1}}$
using \HI 4$\pi$ survey (black dotted line) for a comparison.  As the
spatial resolution is higher for \HI 4$\pi$ Survey, we can probe
larger ${\rm k}$ range $\sim6$ times more than the LAB survey. For the
overlapping range, the two spectra are consistent with each
other. However, the amplitude of the power spectrum for \HI
4$\pi$ falls significantly at ${\rm k}\ge1.5~{\rm deg}^{-1}$.
As shown in Figure 2 of \citet{miville16}, this deviation at
  smaller angular scales is due to the beam shape of the \HI
  4$\pi$ survey data. However, the measured power spectrum
  behaves as a power law for the whole range
  probed by the LAB survey data. It is due to fact that the pixel size
   is roughly similar to the beam size for the LAB survey.
 For the purpose of the present
  analysis, we restrict ourselves only to LAB survey data.The magenta
  points in this figure show the velocity integrated power spectrum in
  the WNM dominated region $(-14 < v_{LSR} <-5 ~{\rm km~s^{-1}})$ for
  \HI 4$\pi$ data. For this case, first, we integrate the image over
  the mentioned velocity range and then calculate the power
  spectrum. We see that the integrated power spectrum remains the same
  as for the single channel power spectrum at $v_{LSR}=-10~{\rm
    km~s^{-1}}$ (discussed later). In terms of the spectral
resolution, LAB survey provides higher resolution data than \HI 4$\pi$
survey. As the spectral resolution is more important for achieving the
main goals of this work, we have used only LAB survey data in the rest
of the analysis.
\begin{figure*}
\begin{center}
\includegraphics[width=55mm,angle=0]{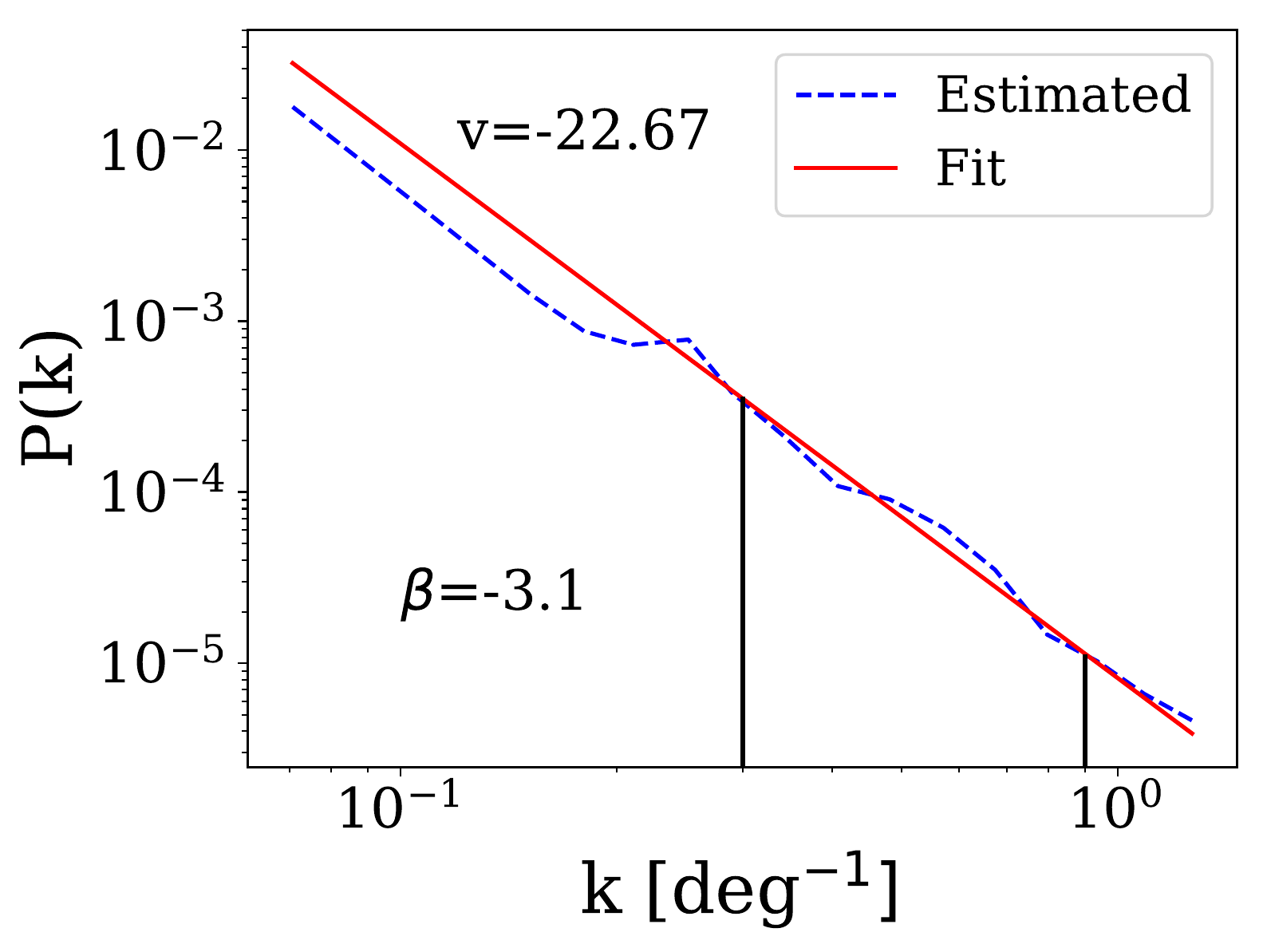}
\includegraphics[width=51mm,angle=0]{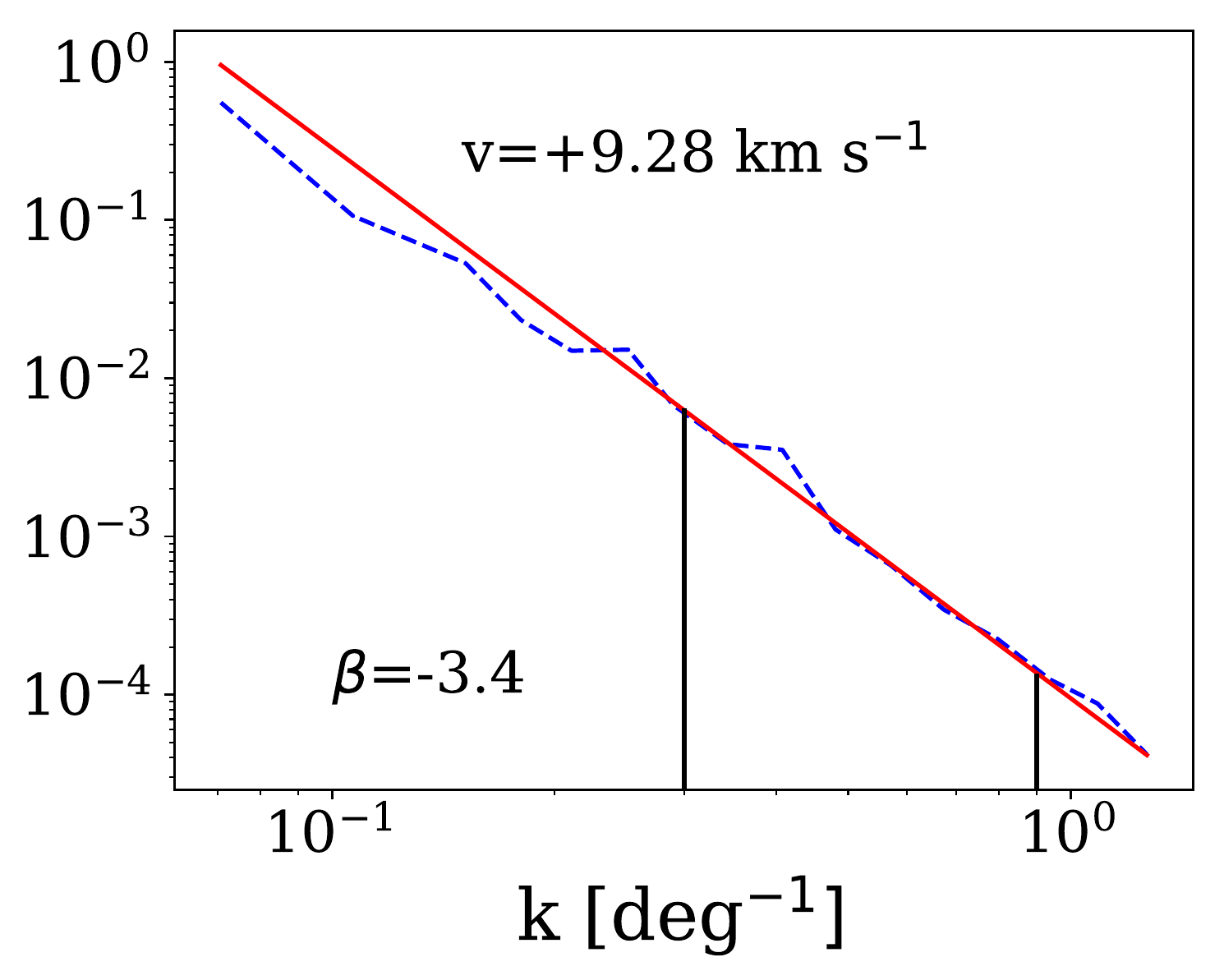}
\includegraphics[width=51mm,angle=0]{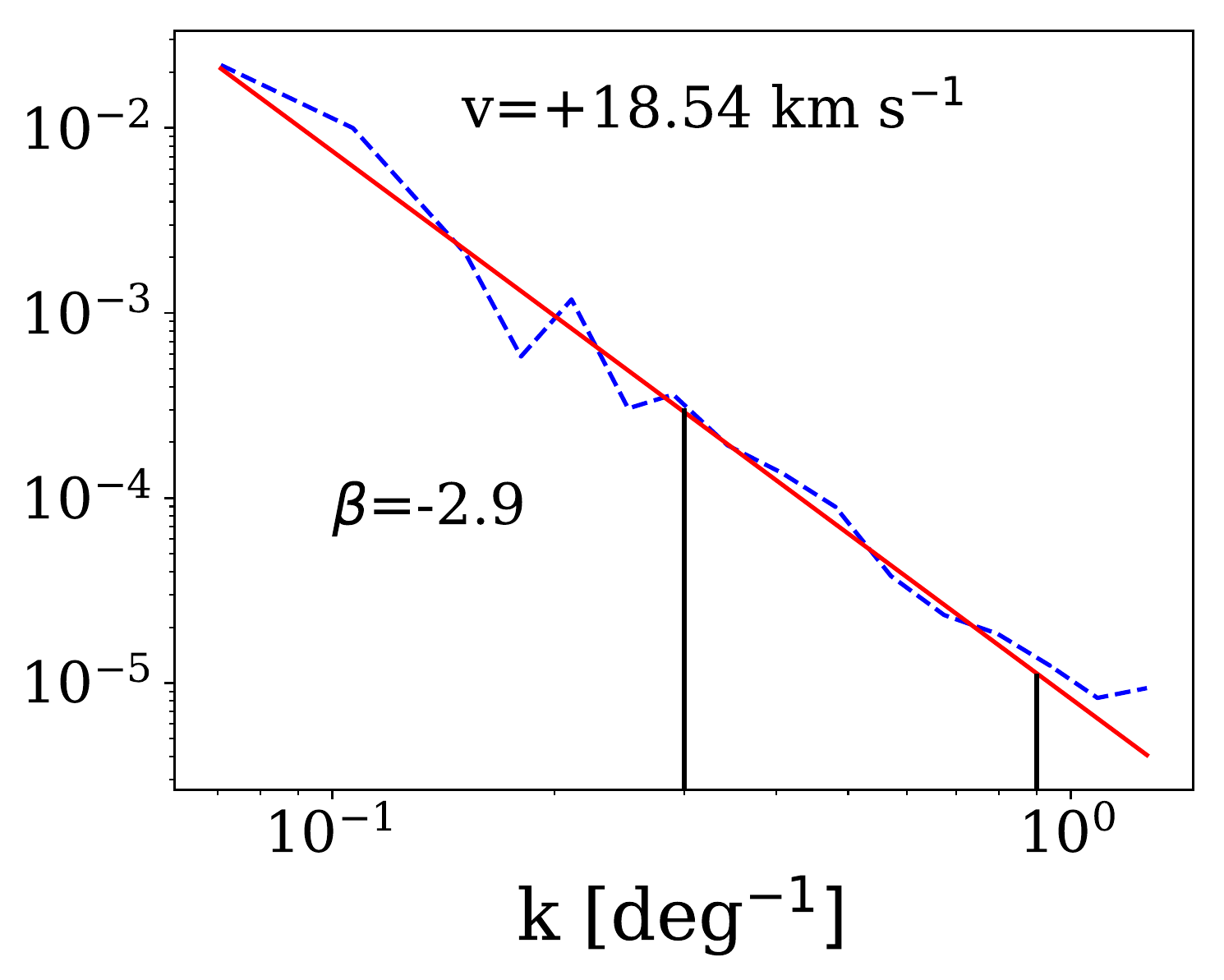}
\caption{The estimated $P(k)$ for three representative 
velocity channels $-22.67$,
  $+9.28$ and $+18.54~{\rm km~s^{-1}}$ which are dominated by IVC, CNM
  and WNM respectively (blue dotted lines). The red lines show the
  best-fit power law model $P^M(k)$. The $k$ range we used for fitting
  is $0.3$ to $0.9~{\rm deg}^{-1}$ which are shown by two vertical
  lines. The values of the power law index $(\beta)$ are shown in the
  lower left corner of each panel.}
\label{fig:fig2}
\end{center}
\end{figure*}
\begin{figure}
\begin{center}
\includegraphics[width=80mm,angle=0]{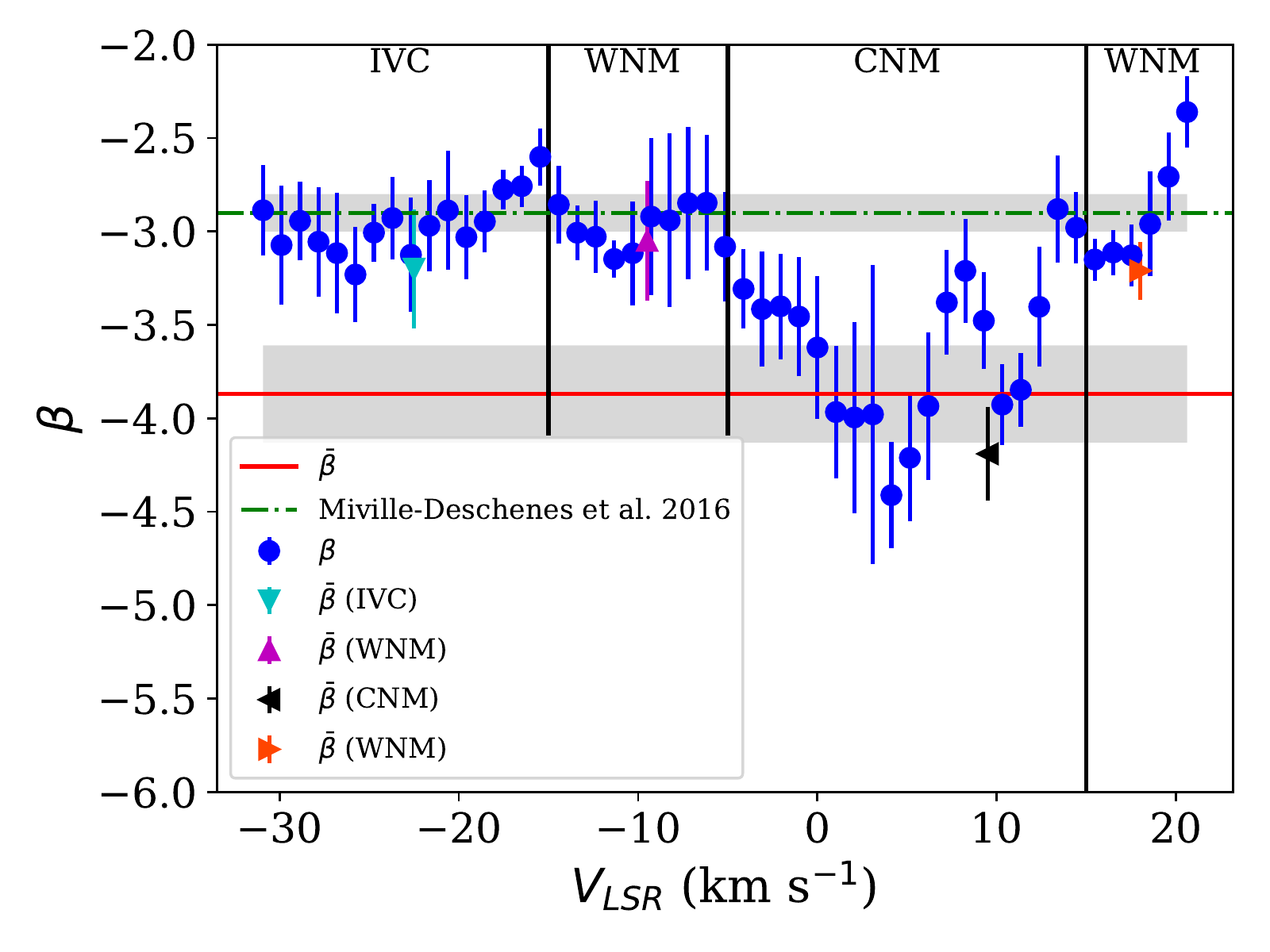}
\caption{The variation of power law index, $\beta$ with $1\sigma$
  error bar as a function of LSR velocity $(v_{LSR})$ (blue
  points). The power-law index, $\bar{\beta}$ estimated from
  integrated column density map for the IVC, WNM (negative velocity
  range), CNM and WNM (positive velocity range) are shown by cyan
  down, magenta up, black left and orange right triangle
  respectively. The green dash-dot line with shaded region is the
  measurement from \citet{miville16}. The red solid line with shaded
  region shows the measured $\bar{\beta}$ estimated from the
  integrated \HI~ emission over the full velocity range of $-30$ to
  $+20~{\rm km~s^{-1}}$.}
\label{fig:fig3}
\end{center}
\end{figure}

Figure \ref{fig:fig2} shows the measured $P({\rm k})$ as a function of
${\rm k}$ for different velocity channels. In this figure, we
  show only three representative  velocity channels $-22.67$, $+9.28$ and $+18.54~{\rm
    km~s^{-1}}$ which are dominated by IVC, CNM and WNM respectively.
  In Figure \ref{fig:fig1a} we show another WNM dominated velocity
  channel from negative velocity range $(-10~{\rm km~s^{-1}})$. The
  blue dashed lines in each panel show the estimated $P({\rm k})$ as a
  function of wavenumber ${\rm k}$.  We fit a power law model
  $P^M({\rm k})=A{\rm k}^{\beta}$ to the measured $P({\rm k})$ and the
  red solid lines show the best-fit power law for each velocity
  channel.  The $k$ range we used for fitting is $0.3$ to $0.9~{\rm
    deg}^{-1}$ which are shown by two vertical lines in this figure.
  The values of the power law index $(\beta)$ are also shown in each
  panel. We see the slope of the CNM dominated region (middle panel)
  is steeper as compared to other regions.

\begin{figure*}
\begin{center}
\includegraphics[width=57mm,angle=0]{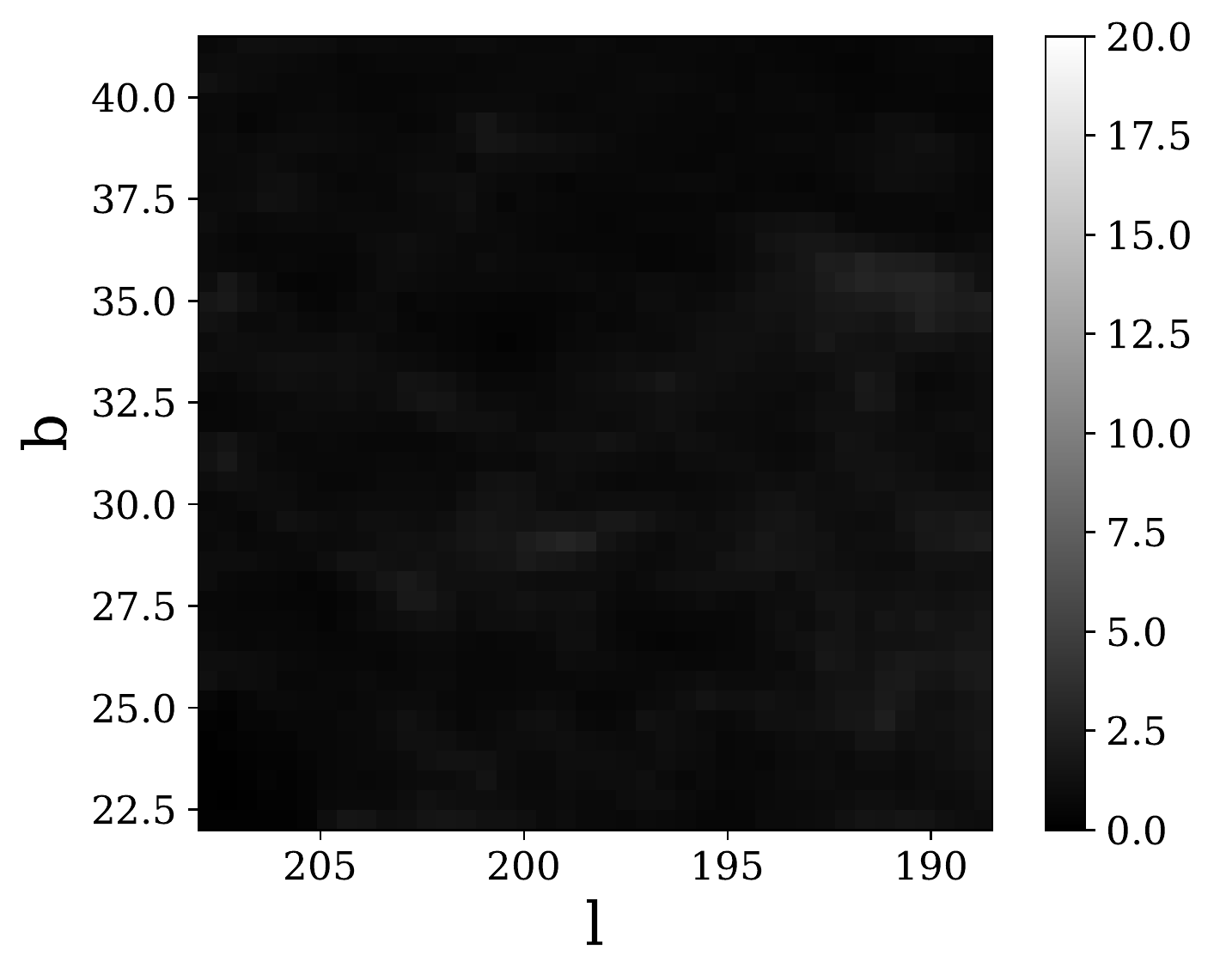}
\put(-70,-5){\large (a)}
\includegraphics[width=54mm,angle=0]{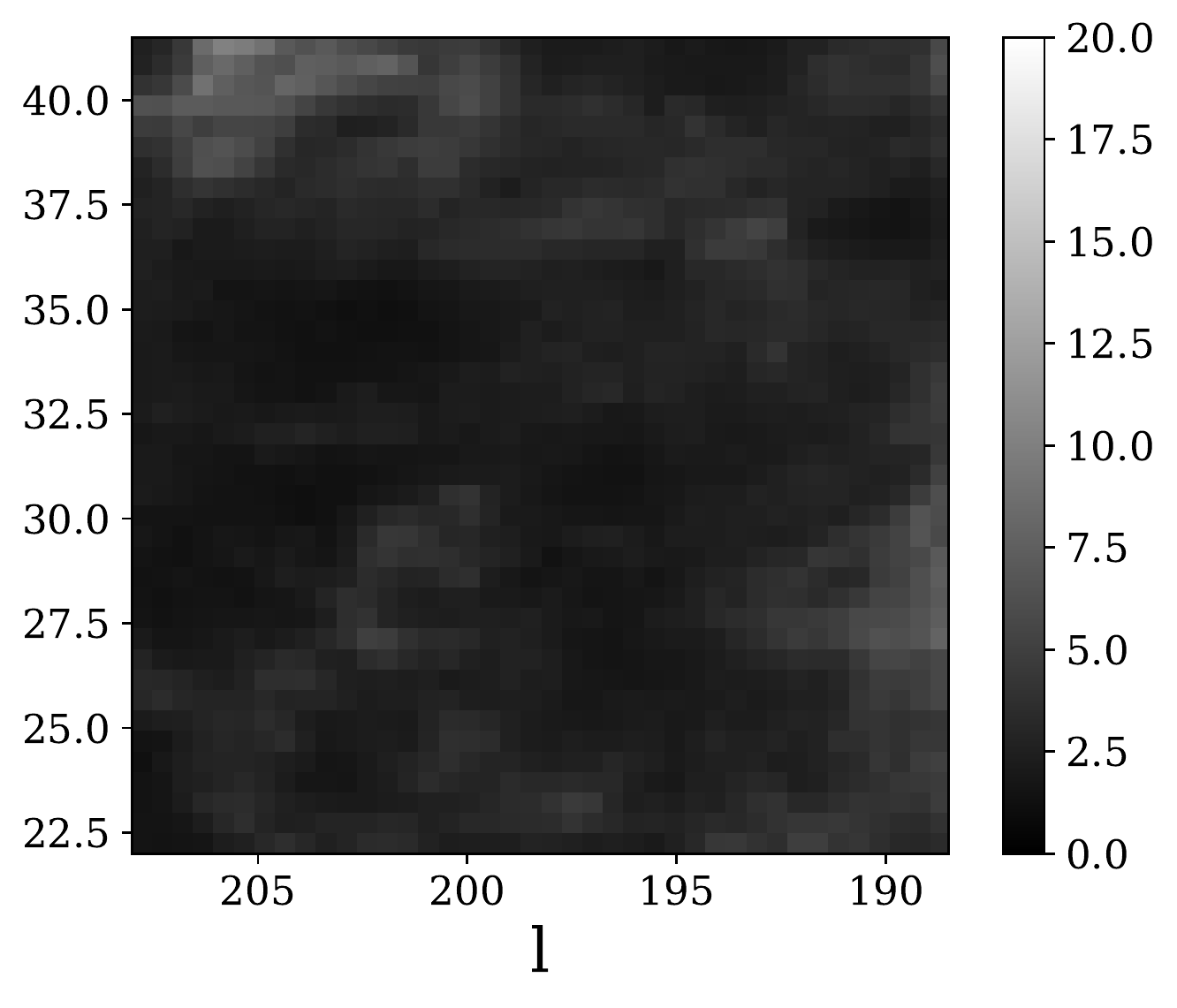}
\put(-70,-5){\large (b)}
\includegraphics[width=54mm,angle=0]{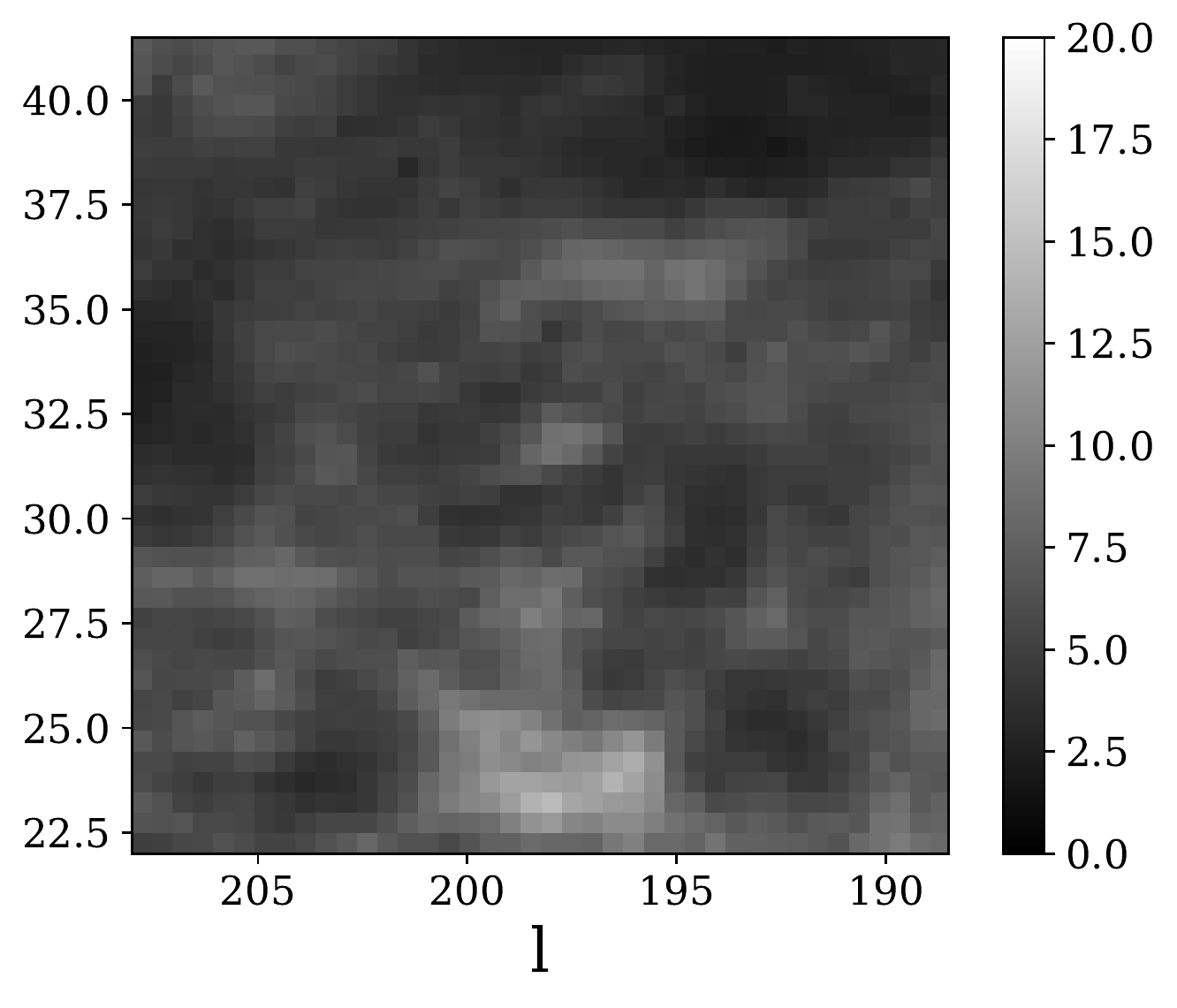}
\put(-70,-5){\large (c)}
\newline
\includegraphics[width=57mm,angle=0]{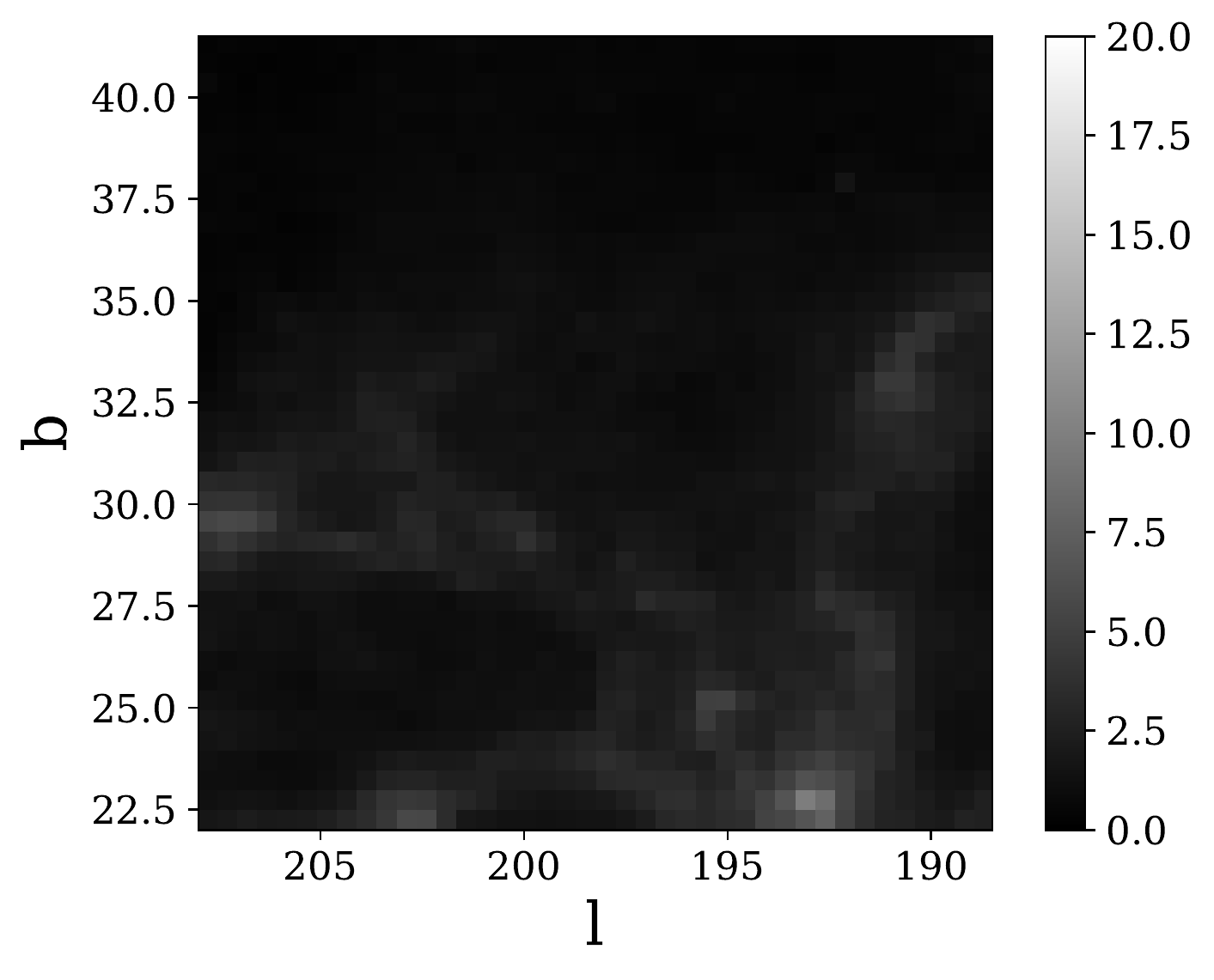}
\put(-70,-5){\large (d)}
\includegraphics[width=54mm,angle=0]{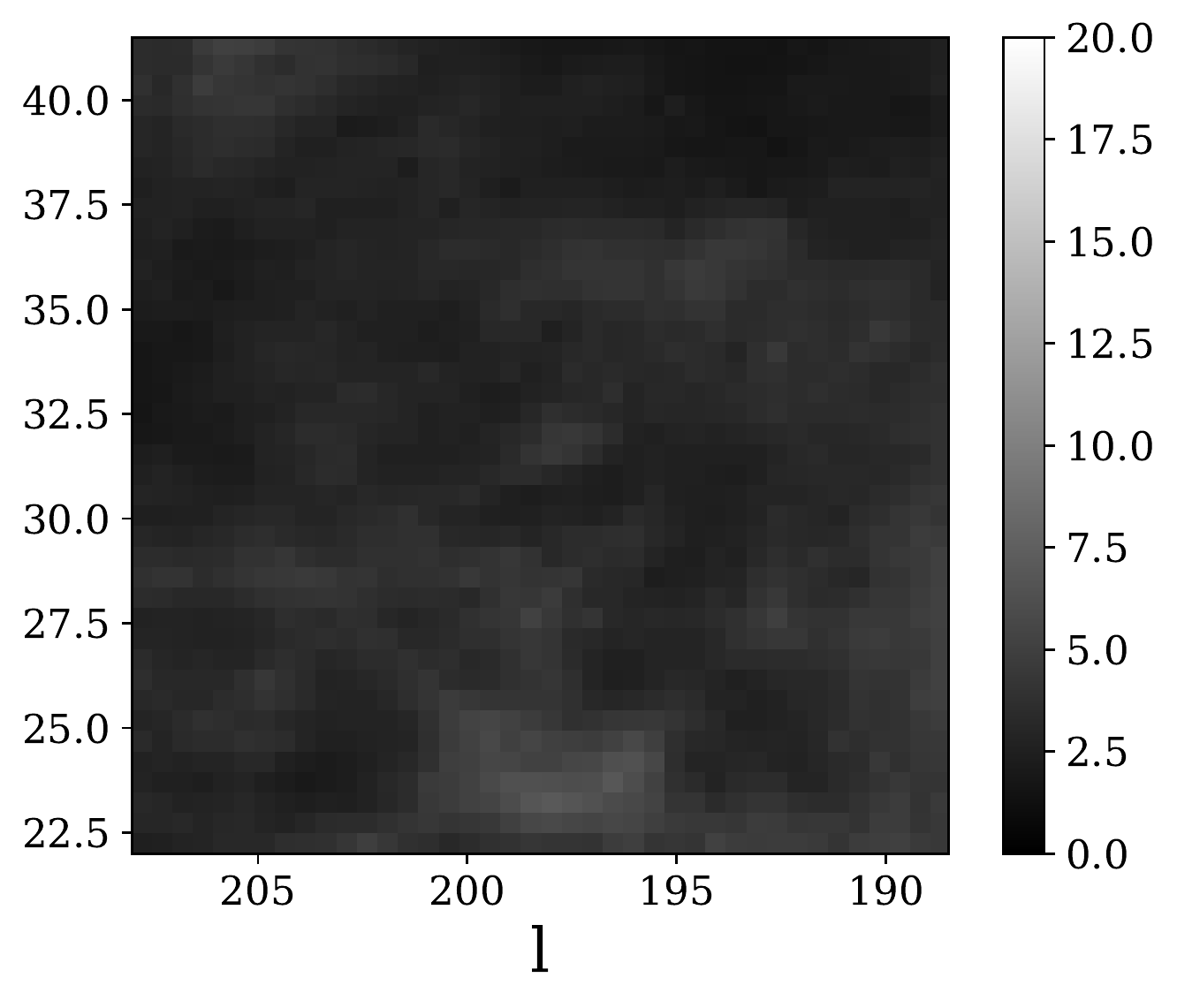}
\put(-70,-5){\large (e)}
\caption{The average \HI~21 cm brightness temperature maps which
    is proportinal to the atomic hydrogen column density $(N_{HI})$ in different
  velocity range: (a) the
  IVC dominated region $(-30 \le v_{LSR} \le -15~{\rm km~s^{-1}})$ (b)
  the WNM (negative velocity range) $(-14 \le v_{LSR} \le -5~{\rm
    km~s^{-1}})$ (c) the CNM $(-4 \le v_{LSR} \le +15~{\rm
    km~s^{-1}})$ (d) the WNM (positive velocity range) $(+16 \le
  v_{LSR} \le +20~{\rm km~s^{-1}})$ (e) the whole velocity range $(-30
  \le v_{LSR} \le +20~{\rm km~s^{-1}})$.}
\label{fig:fig4}
\end{center}
\end{figure*}

We now study the variation of $\beta$ as a function of $v_{LSR}$
(Figure \ref{fig:fig3}). The blue points show the estimated $\beta$
with $1\sigma$ error bar as a function of $v_{LSR}$. The values of
$\beta$ vary between $-2.6$ to $-3.2$ for IVC which dominates in the
velocity range from $-30$ to $-15~{\rm km~s^{-1}}$. The integrated
column density map for this velocity range shown in Figure
\ref{fig:fig4} (a). The integrated column density map is same as
  the average \HI~21 cm brightness temperature map just apart from a
  constant factor. The value of $\bar{\beta}$ estimated using this
integrated map for IVC is $-3.2\pm0.32$ (cyan down triangle in Figure
\ref{fig:fig3}). \citet{martin15} found the power-law index for IVC is
around $-2.69\pm0.04$ which is slightly flatter as compared to our
results. In the WNM dominated region $(-14 < v_{LSR} <-5 ~{\rm
  km~s^{-1}})$ the values of $\beta$ varies in the range $-2.8$ to
$-3.1$ which is quite similar as shown for IVC.  The integrated column
density map for the WNM dominated region is shown in Figure
\ref{fig:fig4} (b).  The slope $\bar{\beta}$ estimated for this image
is $-3.05\pm0.32$ which is (magenta up triangle in Figure
\ref{fig:fig3}) also similar as IVC.

The statistical properties of the CNM which dominate in the velocity
range from $-4$ to $+15~{\rm km~s^{-1}}$ is quite different as shown
for the IVC and WNM. In this case, the $\beta$ values vary in the
range $-2.9$ to $-4.4$. The integrated map for CNM is shown in Figure
\ref{fig:fig4} (c) and corresponding $\bar{\beta}$ is shown by black
left triangle which has a value $-4.19\pm0.25$. Here, the slope is
much steeper as compared to the IVC and CNM. The velocity range $+16$
to $+20~{\rm km~s^{-1}}$ again dominated by the WNM and the integrated
map is shown in Figure \ref{fig:fig4} (d). In this case, the value of
$\bar{\beta}$ is $-3.21\pm0.15$ which is shown by the orange right
triangle in Figure \ref{fig:fig3}). We see that the measured $\beta$
in the CNM dominated velocity region are quite different from that of
the WNM and IVC. This may be indicative of different nature of
turbulence in different thermal phases depending on temperature,
density and ionization fraction.

The power spectrum for the CNM is already measured earlier, and its
power-law index varies in the range $-2$ to $-3$
\citep{kalberla16,kalberla17,blagrave17}. \citet{dickey01} have
measured the power spectrum of the warm HI gas and found power law
index varies in the range $-3$ to $-4$. \citet{miville03} have
estimated the power spectrum of the Ursa Major high-latitude cirrus
and found power law index with a value of $-3.6\pm0.2$. These studies
suggest that the turbulent power spectra cannot be described by a
unique power law.

\begin{figure*}
\begin{center}
\includegraphics[width=165mm,angle=0]{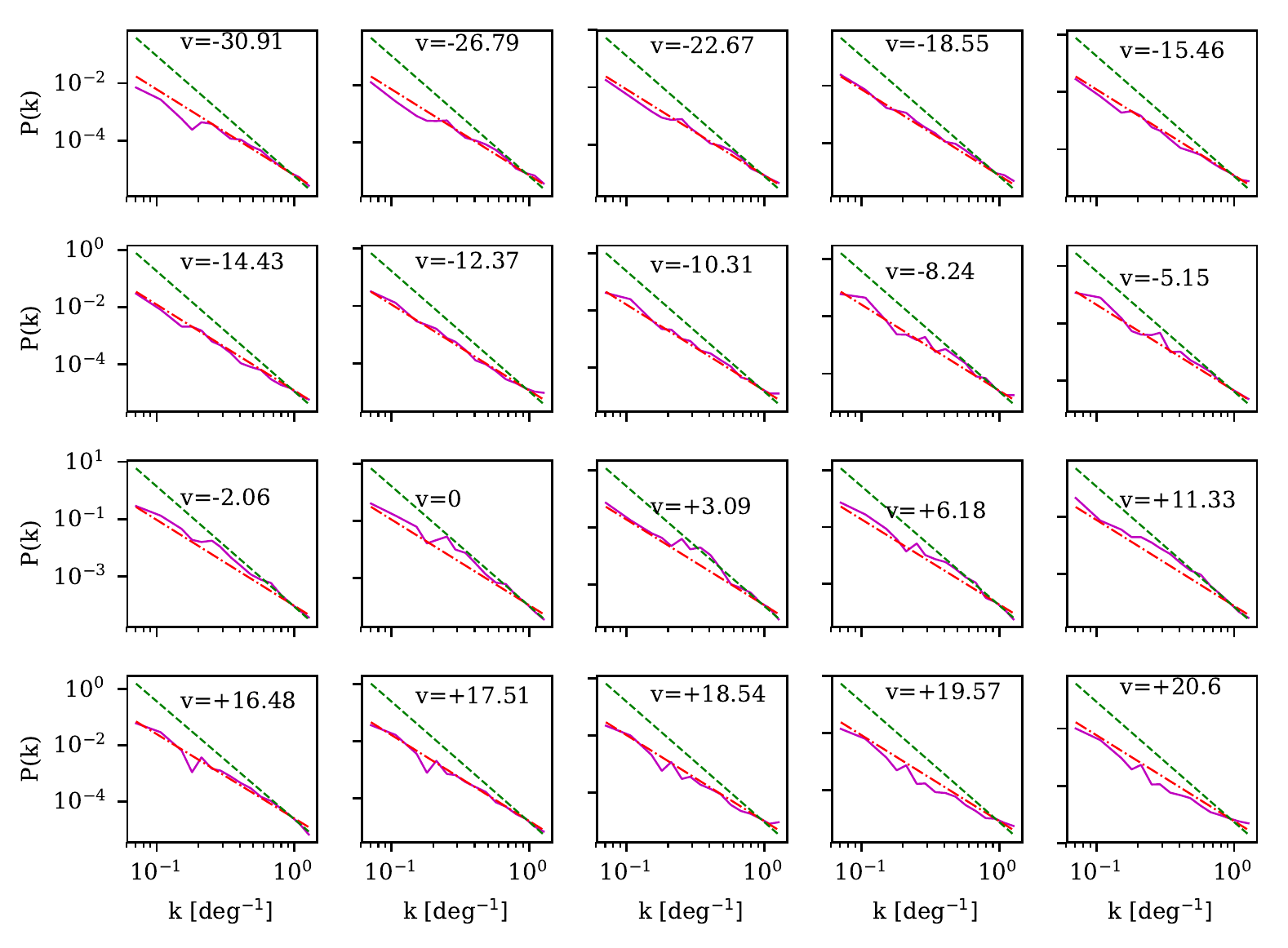}
\caption{The measured 21 cm $P({\rm k})$ (megenta solid line) for 20
  different velocity channels in the range from $-30$ to $+20~{\rm
    km~s^{-1}}$. The green dash lines shows the model $P^M({\rm k})$
  with slope $-3.0$ and the red dash-dot line is for $-4.2$ which are
  the value of $\bar{\beta}$ similar to the WNM (also IVC) and CNM
  dominated region respectively.  Each row shows five different
  velocity channels for differnt components: First row (IVC), Second
  row (WNM in negative velocity range), Third row (CNM), Fourth row
  (WNM in positive velocity range). The power spectrum for the WNM and
  IVC dominated region matches model $P^M({\rm k})$ with $\beta=-3.0$,
  whereas the CNM dominated region matches with $\beta=-4.2$.}
\label{fig:fig5}
\end{center}
\end{figure*}

We have also studied the statistical properties of the integrated
emission for the whole velocity range from $-30$ to $+20~{\rm
  km~s^{-1}}$ and the corresponding image is shown in Figure
\ref{fig:fig4} (e). Using the power spectrum estimated for this image,
we get the best fit value of $\bar{\beta}$ to be $-3.87\pm0.26$ which
is shown by a red solid line with shaded region. Based on dust
continuum emission, \citet{miville16} estimated the value of $\bar{\beta}$
to be $-2.9\pm0.1$ which is shown with a green dash-dot line and a
shaded region in Figure \ref{fig:fig3}. The power law index derived
using the \HI~21 cm emission from the full range of the velocity is
quite different with the value derived by \citet{miville16} for dust
emission.

In Figure \ref{fig:fig5} we show the measured $P({\rm k})$ with solid
magenta lines for 20 velocity channels selected in the range from
$-30$ to $+20~{\rm km~s^{-1}}$. Here, we also show two model $P^M({\rm
  k})$ with power-law index $-3.0$ (red dash-dot line) and $-4.2$
(green dash line) which are the value of $\bar{\beta}$ similar to the WNM
(also IVC) and CNM dominated region respectively. In the first
row, we show five different velocity channels from IVC dominated
regions. Similarly, the second, third and fourth rows are for the WNM
(negative velocity region), CNM and WNM (positive velocity region)
respectively.  We see that the measured $P({\rm k})$ in the IVC and
WNM (both velocity ranges) dominated regions matches the $P^M({\rm
  k})$ with $\bar{\beta}=-3.0$, whereas in the CNM dominated regions
it matches with $\bar{\beta}=-4.2$.

\begin{figure}
\begin{center}
\includegraphics[width=90mm,angle=0]{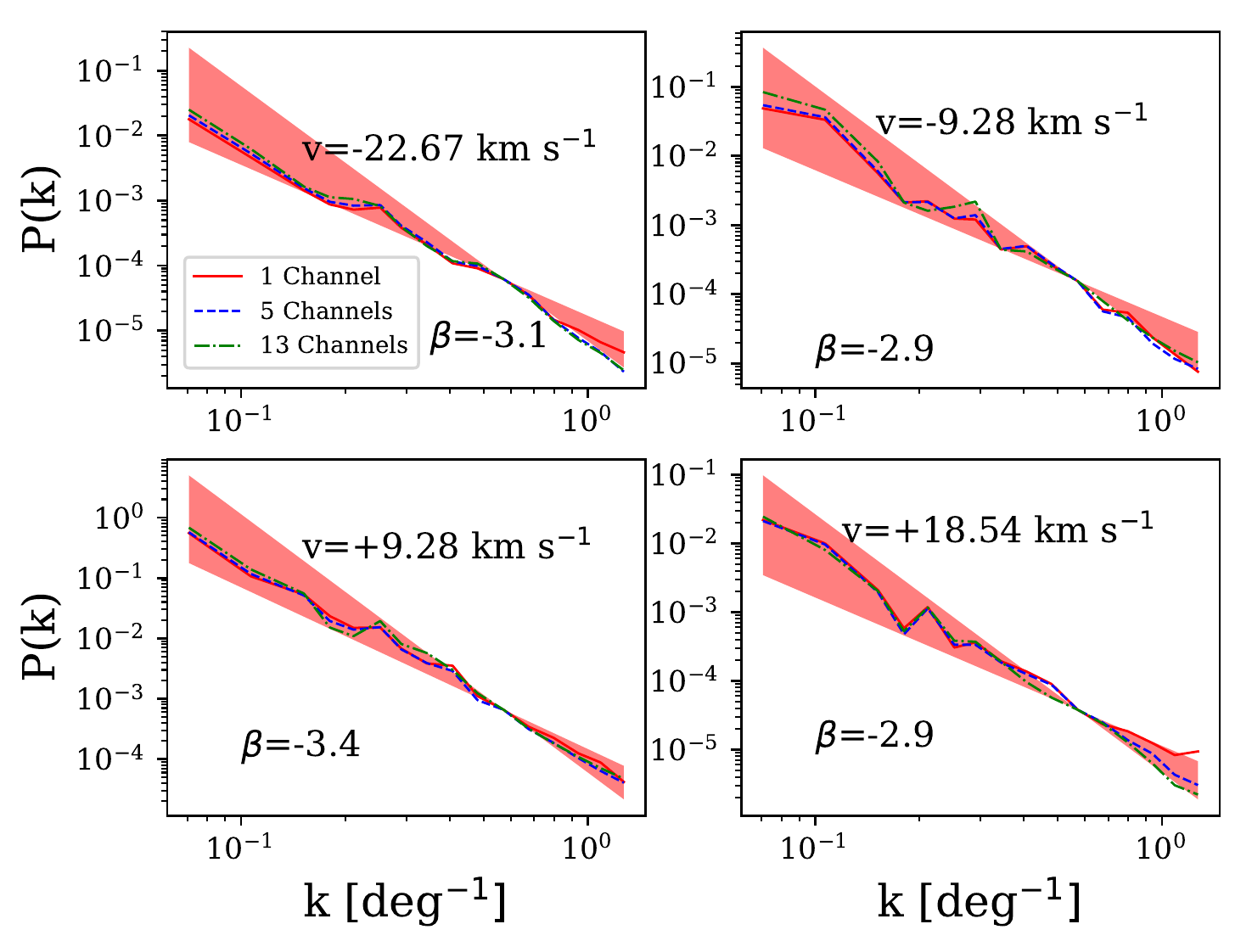}
\caption{The estimated $P({\rm k})$ after averaging different velocity
  channels. The red solid, blue dashed and green dash-dot are for
  $1.03$, $5.15$ and $13.39~{\rm km~s^{-1}}$ respectively for four
  representative velocity channels taken from four different
  regions. The shaded regions show the area for an uncertainty of
  $\pm0.8$ in the power law index $\beta$.}
\label{fig:fig7}
\end{center}
\end{figure}

Next, we have estimated $P({\rm k})$ after averaging the velocity
channels to a width ranging from $1.03$ to $13.09~{\rm
  km~s^{-1}}$. \citet{lazarian00} have shown that the intensity power
spectrum may be changed due to velocity structure in the field. For
``thick'' slices, which are wider than the turbulent velocity
dispersion, all the velocity information get averaged out and the
measured $P({\rm k})$ only quantify the density fluctuations in that
field. However, for ``thin'' velocity slices, the measured $P({\rm
  k})$ becomes shallower due to the velocity structure present in the
field. The power law index of the $P({\rm k})$ for thin slices becomes
$n+\gamma/2$, where $n$ is the power law index for thick slices and
$\gamma$ is the power law index for the velocity structure
  function. The turbulent velocity dispersion for the cold Galactic HI
  gas is around $4.0~{\rm km~s^{-1}}$ \citep{radha72}. The velocity
  width for the CNM we get is almost $5.4~{\rm km~s^{-1}}$ (Figure
  \ref{fig:fig1}) which is slightly higher than the typical velocity
  dispersion of cold gas. In this study, we considered the velocity
  width from $1.03$ to $13.09~{\rm km~s^{-1}}$. We can not probe the
  structures of velocity width below $1.25~{\rm km~s^{-1}}$ as we are
  limited by the resolution of the LAB survey.  We expect
  the change in the power law index for thin slices of a width smaller
  than velocity dispersion. In Figure \ref{fig:fig7} we show the
measured $P(k)$ only for three velocity width $1.03$ (red solid line)
$5.15$ (blue dashed line) and $13.39~{\rm km~s^{-1}}$ (green dash-dot
line) for clear visualization. Also, we show only four representative
velocity channels ($v_{LSR}=-22.67, -9.28, +9.28$ and $18.54~{\rm
  km~s^{-1}}$) in this figure, even if the exercise is repeated for
all channels (and the results are the same for all the channels). We
choose these four velocity channels from four different
regions. Clearly, we do not observe any significant difference in the
power spectrum after velocity channel averaging. This signifies that
the fluctuations are mainly dominated by the density
fluctuations. We use the maximum error in measured $\beta$
$(\sim0.8)$ to constrain the value of $\gamma$ to be $0.0\pm1.1$
which is consistent with the predicted Kolmogorov
turbulence $(\gamma=2/3)$ and also with a
    shock-dominated medium $(\gamma=1.0)$.

\section{Summary and Conclusions}
\label{summa}
Measurements of \HI~21 cm power spectrum, $P({\rm k})$ can be used to
quantify the small-scale structures within our Galaxy. In this paper,
we have studied the 21 cm $P({\rm k})$ over a length scale ranging
from $3~{\rm pc}$ to $50~{\rm pc}$. We have used high-resolution LAB
survey data to measure the $P({\rm k})$ of a
$20^{\circ}\times20^{\circ}$ region centred at the Galactic coordinate
$(198^{\circ},32^{\circ})$. Here, we have considered a velocity range of
$-30$ to $+20~{\rm km~s^{-1}}$ which we believe to be dominated by 21
cm signal only. The non-detection of molecular $^{12}CO(J = 1 - 0)$
  emission in this field \citep{miville16} also indicates the
  same. \citet{miville16} have already measured the $P({\rm k})$ of
continuum dust emission in this region using the emission maps from
three different surveys. They found $P({\rm k})$ follows a power law
with index $-2.9\pm0.1$. We see that the measured 21 cm $P({\rm k})$
also follows a power law within the scale range probed here.  However,
the power law index was found to have a significant channel to channel
variation over this velocity range. Also, the best fit value of
  the power-law index $\bar{\beta}$ estimated from the integrated map
  from different regions in this velocity ranges are quite different.
  The values of $\bar{\beta}$ are
  $-3.2\pm0.32,-3.05\pm0.32,-4.19\pm0.25$ and $-3.21\pm0.15$ for the
  IVC, WNM (negative velocity range), CNM and WNM (positive velocity
  range) dominated regions respectively. We see that the measured
  $\bar{\beta}$ in the CNM dominated region is significantly different
  as compared to other components e.g. IVC, WNM.  A possible
  interpretation of this can be an inherent quantitative difference of
  the turbulence power spectra for different phases of the ISM.  There
  is no a priori reason to assume that the nature of turbulence
  identical in these phases.  In the absence of such universality, the
  measured $P({\rm k})$ may vary depending on the relative
  contribution of these phases.

Please note that, while modelling the observed power spectrum, we
assume a power law form based on the fact that small-scale structures
generated due to turbulence will have such scaling between the
injection scale and the dissipation scale. For Kolmogorov-like
turbulence in the neutral gas, the Reynold's number is given by $R_{e}
= 3\times 10^{4}\, \frac{L}{10\, {\rm pc}}\, \frac{v}{10\, {\rm
    km~s}^{-1}}\, \frac{n}{1\, {\rm cm^{-3}}}$ and the dissipation
scale is $l_{d} = L R_{e}^{-3/4}$ where $v$ is the turbulent velocity
dispersion, $n$ is the average density at the energy dissipation
scale, and $L$ is the energy injection scale for turbulence
\citep{ks98}.  If we consider the two components with $\sigma_v = 5.4$
and $17.3~{\rm km~s^{-1}}$ to be cold and warm neutral medium
respectively, with typical parameters like $T_k = 100$ and $5000~{\rm
  K}$, and density in the range of $10 - 100$ and $0.1 - 1~{\rm
  cm^{-3}}$ for these two phases, the dissipation scales turn out to
be $2.23\times 10^{-4}$ - $1.25\times 10^{-3}~{\rm pc}$ for CNM and
$3.07\times 10^{-3}$ - $1.73\times 10^{-2}~{\rm pc}$ for WNM, well
below the $3.4~{\rm pc}$ scale probed here using LAB data.

We have also studied the power spectrum after averaging the velocity
of width ranging from $1.03$ to $13.39~{\rm km~s^{-1}}$.  For thin
velocity slices, which are smaller than the turbulent velocity
dispersion $(4.0~{\rm km~s^{-1}})$, the measured $P(k)$ becomes
shallower due to the velocity fluctuations present in the field
\citep{lazarian00}. However, we do not observe any significant
difference in measured $P({\rm k})$ after velocity channel averaging.
This implies that either the transition happens below $1.25~{\rm
    km~s^{-1}}$ which is the resolution of the LAB survey, or the
  velocity structure function is shallower than our measurement error
  of the power law index of the power spectra. From this study, we can
  only constrain the value of $\gamma$ to be $0.0\pm1.1$ which is
  consistent with the predicted Kolmogorov turbulence $(\gamma=2/3)$
  and also with a
    shock-dominated medium $(\gamma=1.0)$.
We plan to investigate this issue with better spectral resolution data
in future.

Finally, similar high-velocity resolution investigation of Galactic
\HI~power spectra for different directions will be helpful to
understand if the nature of turbulence is truly different for the cold
and warm phases.

\section{Acknowledgements}
We thank an anonymous referee for helpful comments. 
SC acknowledges NCRA-TIFR for providing
financial support. SC would like to thank Narendranath Patra for
useful discussion. SC acknowledges NR's Infosys Young Investigator
grant for supporting his collaborative visit to IISc and IISc for an
opportunity to work on this project during the visit. NR acknowledges
support from the Infosys Foundation through the Infosys Young
Investigator grant.

\end{document}